\documentstyle[12pt]{article}

\def\vecmu{\mbox{\boldmath $\mu$}}
\def\vecrho{\mbox{\boldmath $\rho$}}

\begin{document}

\centerline{$E \otimes \varepsilon$ Jahn-Teller Anharmonic Coupling
for an Octahedral System}

\bigskip

\centerline{N.M. Avram,\footnote{Permanent address: Department of
Physics, University of the West Timi\c soara, Bd V. P\^ arvan Nr. 4, 
1900 Timi\c soara, Romania} 
Gh.E. Dr\v ag\v anescu\footnote{Permanent address: Department of
Physics, University of the West Timi\c soara, Bd V. P\^ arvan Nr. 4, 
1900 Timi\c soara, Romania} 
and M.R. Kibler\footnote{To whom correspondence should be addressed.}}

\medskip

\centerline{Institut de Physique Nucl\' eaire de Lyon}

\centerline{IN2P3-CNRS et Universit\' e Claude Bernard}

\centerline{43, Bd du 11 Novembre 1918}

\centerline{F-69622 Villeurbanne Cedex, France}

\bigskip

\begin{abstract}
The coupling between  doubly degenerate 
electronic states and doubly degenerate vibrations
is analysed for an octahedral system on the basis 
of the introduction of an anharmonic Morse potential 
for the vibronic part. The
vibrations are described by anharmonic coherent states and their  
linear coupling with the electronic states is considered. The matrix
elements of the vibronic interaction are builded and the energy levels 
corresponding to the interaction Hamiltonian are derived.
\end{abstract}

\section{Introduction}

The case of an octahedrally coordinated metal-ion
presenting a doubly degenerate (non-Kramers) electronic 
state of symmetry $E$ interacting with a doubly 
degenerate vibrational state of symmetry 
$\varepsilon$ constitutes one of the most studied 
Jahn-Teller system [1-9]. In a cluster model \cite{010}, 
the Hamiltonian $H$ of the system reads
\begin{equation}
H = H_e + H_v + H_{JT}.
\label{efg1}
\end{equation}
In Eq.~(\ref{efg1}), $H_e$ stands for the Hamiltonian of the
electronic part, $H_v$ represents the vibrational
Hamiltonian and $H_{JT}$ is the Jahn-Teller interaction. To fix the
notations, we briefly review the different contributions to $H$ in 
the usual case of harmonic vibrations. 

The electronic Hamiltonian $H_e$ is well-known. For transition metal-ions,
it is convenient to work in the strong-field coupling scheme. We denote as
$| \theta \rangle$ and $| \varepsilon \rangle$ the uncoupled electronic
wavefunctions corresponding to $H_e$ and 
transforming according to the two-dimensional irreducible 
representation $E$ (or $E_g$) of the octahedral (or complete octahedral) 
point group $O$ (or $O_h$).

We consider vibrations for the doubly degenerate modes 
with normal coordinates $Q_1$ and $Q_2$ sharing a 
common frequency $\omega$. Let $P_1$ and $P_2$
be the generalized momenta associated to $Q_1$ and 
$Q_2$, respectively. Both sets 
$\{ Q_1 , Q_2 \}$ and
$\{ P_1 , P_2 \}$ transform as basis
functions for the irreducible representation $E$ of  
the group $O$. The vibrational Hamiltonian $H_v$ is
  \begin{equation}
H_v = {1          \over 2 m} (P_1^2 + P_2^2) 
    + {m \omega^2 \over 2  } (Q_1^2 + Q_2^2), 
  \label{e2}
  \end{equation}
where $m$ is the effective mass for the two-dimensional harmonic
oscillator in the plane $( Q_1, Q_2 )$.
The operator $H_{v}$ can be expressed as
$$                                                 
H_v = \hbar \omega (a_1^{\dag} a_1 +
                    a_2^{\dag} a_2 + 1)
$$
in terms of $\hbar \omega$ 
and of the creation operators $a^{\dag}_k$ and the
            annihilation operators $a_k$ associated to the 
couples ($Q_{k}, P_{k}$) with $k = 1,2$. 

The coupling between the harmonic vibrations of the ligands and the
electronic functions localized around the metal-cation can
be described by means of the linear Jahn-Teller Hamiltonian
\begin{equation}
H_{JT} = \sqrt{2 m \omega^2 E_{JT}} (Q_1 D_\theta + Q_2 D_\varepsilon).
\label{e456}
\end{equation}
In Eq.~(\ref{e456}), $D_\theta$ and $D_\varepsilon$ are normalized
operators in the electronic space with the well-known representation
[11,12]
$$                                                 
D_\theta = \left( 
\begin{array}{c c}                                         
-1  &0\\
 0  &1\\
\end{array}
\right), \quad
D_\varepsilon =
\left(
\begin{array}{c c}
0 &1\\
1 &0\\
\end{array}
\right)
$$
and $E_{JT}$ is the Jahn-Teller energy. The energy $E_{JT}$ 
is connected to the vibronic coupling constant
$K$ by the relationships
$$                                                 
K = \sqrt{\hbar \omega E_{JT}}.
$$
The interaction Hamiltonian $H_{JT}$ can thus be rewritten as
$$                                                 
H_{JT} = K [(a_1^{\dag} + a_1) D_\theta 
       +    (a_2^{\dag} + a_2) D_\varepsilon]
$$
in terms of $K$ and of the 
operators $a^{\dag}_k$ and $a_k$ with $k = 1,2$. A simple 
consideration 
of the interaction Hamiltonian $H_{JT}$ shows that the 
mode $Q_1$  
splits the two-fold degeneracy of the electronic partners
$| \theta \rangle$ and $| \varepsilon \rangle$ while 
the mode $Q_2$ mode mixes 
$| \theta \rangle$ and $| \varepsilon \rangle$. 

In order to obtain analytic expressions for quantities of
physical interest, as for example the energies of the 
levels or the Ham factor,  it  is
necessary to use wavefunctions for the part $H_e + H_v$
of the Hamiltonian (\ref{efg1}). Such wavefunctions are products of 
electronic wavefunctions 
(of type $| \theta \rangle$ or $| \varepsilon \rangle$)
with wavefunctions that describe the vibration of the ligands.
For the latter wavefunctions, we can use coherent states 
arising from an
harmonic oscillator with a linear potential that changes the
equilibrium point but not the frequency. 

The use of the coherent
states in the Jahn-Teller effect goes back  to  the idea that the
coupling between local vibrations and electronic orbitals involves 
an operator linear in the normal modes in a first approximation
[13,14]. In this direction, the study of the Jahn-Teller effect 
using Glauber coherent 
states as well as some applications to particular systems are well-known
[15-19]. The Glauber coherent states can be written as
\begin{equation}                                                 
| G(\vecrho) \rangle = \exp \left( - \frac{1}{2} {|\vecrho|^2} \right) 
\exp(\vecrho { \bf a^{\dag} }) | 00 \rangle ,
\label{e8910}
\end{equation}
where $|00 \rangle $ is the usual ground state for the two-dimensional
harmonic oscillator with vanishing occupation numbers $n_1$
and $n_2$. In Eq.~(\ref{e8910}), we use the
vectors $\vecrho = (\rho_1, \rho_2)$
and     ${\bf a^{\dag} } = ( a^{\dag}_1, a^{\dag}_2)$.
In view of 
$$     
a_k | G(\vecrho) \rangle = \rho_k | G(\vecrho) \rangle,
$$
the Glauber coherent state $| G(\vecrho) \rangle$ is an
eigenstate of 
the annihilation operator $a_k$ with the complex 
eigenvalue $\rho_k$ ($k = 1, 2$).

In orbital doublet systems, the potential energy surface, 
in the linear coupling approximation, has a continuous set of minimas which 
are commonly described as forming a "Mexican hat". In 
actual (i.e., physical) $E \otimes \varepsilon$ systems, isolated 
minimas arise 
either from an ion~-~lattice interaction term which is quadratic in the
displacements 
of the normal coordinates or 
from vibrations which are anharmonic in such 
coordinates. Both terms warp the potential energy surface so 
that three minimas are formed as 
a result of one of these terms acting either separately or together
with the others. 
For these reasons the surface that represents the adiabatic potential 
warps and, along the bottom of the trough of the hat, three wells occur, 
alternating regularly with three humps known as "tricorns" [8]. 
        \" Opik and Pryce [2] and O'Brien [6] took into account the 
quadratic terms of vibronic coupling in the Jahn-Teller Hamiltonian and 
established the influence of these terms on the energy levels of the 
system.

In general, the replacement of the harmonic oscillator potential by
the anharmonic Morse potential [20,21] turns out to be an efficient way
for describing anharmonicity of the vibrations. On the other hand, the 
anharmonic coherent states for the Morse potential have been 
recently investigated with great details \cite{10}. Therefore, it
is the aim of the present paper to
consider a deviation from the harmonic 
approximation in the nuclear motions, i.e., to 
explicitly introduce the vibronic anharmonicity
by using the 
Morse potential. More precisely, we 
study in this work the case of an octahedral Jahn-Teller
system for which the doubly degenerate electronic states are coupled
with the doubly degenerate vibrational states
taken as the anharmonic coherent states for a 
two-dimensional Morse potential.

\section{Morse anharmonic vibrations}

The Hamiltonian $H$ of the $E \otimes \varepsilon$ octahedral Jahn-Teller
system has the same form as in (\ref{efg1}) but  the contribution  $H_v$ 
represents now  the vibrational Hamiltonian  for  a 
two-dimensional Morse oscillator. We thus replace the Hamiltonian
(\ref{e2}) by the double Morse oscillator Hamiltonian:
   $$
H_v = \sum_{k = 1}^2 H_{v_k} 
   $$
with 
$$
H_{v_k} = {p_k^2 \over 2 m} 
        + V_0 ({\rm e}^{- 2 \alpha x_k} - 2 {\rm e}^{- \alpha x_k}), 
$$
where $m$ is as before the effective mass for the oscillator, 
$\alpha$ (with $\alpha > 0$) the anharmonic constant
and $V_0$ a positive constant. Furthermore, $x_k$ is the
displacement from the equilibrium position for the $k$-th dimension
and $p_k$ the associated momentum ($k=1,2$). (The new conjugated 
variables $x_k$ and $p_k$ correspond to the 
variables $Q_k$ and $P_k$ of Section 1, respectively.) We also use
the reduced parameter 
$$
\nu = \sqrt{8 m V_0 \over \alpha^2 \hbar^2}.
$$

The eigenvalues $E_0(n_k)$ of the operator $H_{v_k}$ are given by
$$
E_0(n_k) = - \hbar \Omega \left( n_k - {{\nu - 1} \over 2} \right)^2,
$$ 
where 
$$
\hbar \Omega = 4 \frac{V_0}{\nu^2}
$$ 
and 
$$
n_k = 0, 1, \cdots, N
$$
with
$$
N = \left[ {\nu -1} \over 2 \right].
$$
(As usual, $[r]$ denotes the entire part of the real number $r$.)

The eigenfunction $\psi_{n_k} : {\bf R}_+ \to {\bf R}$ of $H_{v_k}$
corresponding to the level $E_0(n_k)$ is given by \cite{8}
\begin{equation}
\psi_{n_k} ( y_k ) = c_{n_k} y_k^{s_k} {\rm e}^{ - {1 \over 2} y_k } 
                     F (- n_k, 2 s_k + 1, y_k  ), 
\label{treize}                  
\end{equation} 
in terms of the new variable
$$
y_k = \nu {\rm e}^{- \alpha x_k}.
$$ 
In Eq.~(\ref{treize}), the function $F$ is the confluent hypergeometric
function,
the parameter $s_k$ reads
$$
s_k = {{\nu - 1} \over 2} - n_k,
$$
and the factor
$$
c_{n_k} = {1 \over \Gamma (\nu - 2 n_k)} \sqrt{ \Gamma (\nu - n_k) \over n_k !} 
$$
is a normalization constant. 
Finally, we shall use the Dirac notation 
$$
| n_1,n_2 \rangle = \psi_{n_1} ( y_1 ) \psi_{n_2} ( y_2 )
$$
for the vibrational
eigenstates of the two-dimensional Morse oscillator. In this
notation, the relationships
$$
\langle n_1,n_2 | n_1' , n_2' \rangle = \delta (n_1' , n_1) \delta (n_2' ,
n_2).
$$ 
expresses the orthonormalization of the states $| n_1,n_2 \rangle$ 
on the Hilbert space $L^2({\bf R}_+^2, dx_1 dx_2)$.

\section{The anharmonic coherent states}
 
In some previous works \cite{10},  we  have  discussed  
the r\^ole played by  a creation     operator $b_+$,
                     an annihilation operator $b_-$  and  an  
                      energy         operator $b_0$
for describing the dynamical algebra of the one-dimensional 
Morse oscillator. The results of
Ref.~\cite{10} can  be extended to a  two-dimensional isotropic 
Morse oscillator by
introducing creation operators $b_{+ k}$, annihilation
operators $b_{- k}$ and energy operators $b_{0 k}$ (with $k = 1, 2$)
that parallel the operators $b_+$, $b_-$ and $b_0$, respectively. 
The operators $b_{{\pm}k}$ and $b_{0k}$ are defined by
the differential forms
\begin{equation}
b_{\pm k} = (2 s_k \mp 1) { \partial\over \partial y_k} \pm { s_k
(2 s_k \mp 1)\over y_k } \mp {\nu \over 2},
\label{fgh11}
\end{equation} 
\begin{equation}
b_{0 k} = - y_k {\partial^2\over \partial y_k^2} - 
{\partial\over
\partial y_k} + {s_k^2\over y_k} + {y_k\over 2}
- s_k + {\nu\over 2} - 1, 
\label{fgh12}
\end{equation} 
with $k =1,2$. They obey the commutation relations
$$
[ b_{+ k}   , b_{- l} ] = 2 b_{0 k}     \delta (k , l), \quad
[ b_{\pm k} , b_{0 l} ] = \pm b_{\pm k} \delta (k , l),
$$
where $k$ and  $l = 1 , 2$. Therefore, the set 
$\{ b_{\pm k} , b_{0k} : k = 1,2 \}$ generates the Lie algebra of the group 
${\rm SU}(1,1) \otimes {\rm SU}(1,1)$. The action of this set 
on the space $L^2({\bf R}_+^2, dx_1 dx_2)$ is defined through
$$
b_{+ 1} | n_1,n_2 \rangle = \sqrt{(n_1 + 1)(\nu - n_1 - 1)} |n_1 + 1 , n_2
\rangle,
$$
$$
b_{+ 2} | n_1,n_2 \rangle = \sqrt{(n_2 + 1)(\nu - n_2 - 1)} |n_1 , n_2 + 1
\rangle,
$$
$$
b_{- 1} | n_1,n_2 \rangle = - \sqrt{n_1 (\nu - n_1)} |n_1 - 1 , n_2 \rangle,
$$
$$
b_{- 2} | n_1,n_2 \rangle = - \sqrt{n_2 (\nu - n_2)} |n_1 , n_2 - 1 \rangle,
$$
$$
b_{0 k} | n_1,n_2 \rangle = ({\nu - 1\over 2} - n_k) | n_1,n_2 \rangle, \quad k
= 1,2.
$$

The operators $b_{{\pm}k}$ and $b_{0k}$ given 
by (\ref{fgh11}) and (\ref{fgh12})  are  
$s_k$-dependent, i.e., energy-dependent. It is hence necessary to 
introduce new generators of the Lie group ${\rm SU}(1,1) \otimes {\rm SU}(1,1)$
which do not depend on  $s_k$  and 
which have an action on the states $| n_1 , n_2 \rangle$ similar to the one of 
$b_{{\pm}k}$ and $b_{0k}$. This may be achieved with the aid of two auxiliary
variables
$\xi_k$ of a phase type ($\xi_k \in [0, 2 \pi]$ for $k = 1, 2$). To be precise,
let us
define the operators
$$
a_{\pm k} = {\rm e}^{\mp  {\rm i} \xi_k} \left\{ \left[ {2 \over {\rm i}}
{\partial\over \partial \xi_k} \mp 1 \right]  {\partial \over
\partial y_k} \pm {1\over y_k} \left[ {1 \over {\rm i}} {\bf \partial
\over \partial \xi_k } \left( {2 \over {\rm i}} {\partial \over \partial
\xi_k } \mp 1 \right) \right] \mp {\nu \over 2} \right\}
$$
and
$$
a_{0 k} = {1 \over {\rm i}} {\partial\over \partial \xi_k},
$$
with $k = 1,2$. The commutators of the operators $a_{+k}$, $a_{-k}$ and
$a_{0k}$ 
are
$$
[ a_{+ k}   , a_{- l} ] = 2 a_{0 k}     \delta (k , l), \quad
[ a_{\pm k} , a_{0 l} ] = \pm a_{\pm k} \delta (k , l),
$$
with $k$ and $l = 1,2$. In addition, we have 
$$
[ a_{0 k}, {\rm e}^{\pm {\rm i} \xi_l}] = \pm 
           {\rm e}^{\pm {\rm i} \xi_k} \delta (k , l),
$$
with $k$ and $l = 1, 2$.

As a result of our transformation from the $b$'s to the $a$'s, 
it appears that the wave-function 
$\psi_{n_k}$ belongs to a ray defined via
$$
\Phi_{n_k} ( y_k, \xi_k ) = {\rm e}^{ {\rm i} s_k \xi_k } \psi_{n_k} ( y_k ).
$$
The function $\Phi_{n_k}$ is of course an eigenfunction of $H_{v_k}$
with the eigenvalue $E_0(n_k)$ and the action of  
the operators $a_{\pm k}$ and $a_{0 k}$ on $\Phi_{n_k}$ is
$$
a_{+ l} \Phi_{n_k}     ( y_k, \xi_k ) = \sqrt{(n_k + 1)(\nu - n_k -1)}
        \Phi_{n_k + 1} ( y_k, \xi_k ) \delta (k , l),
$$
$$
a_{- l} \Phi_{n_k}     ( y_k, \xi_k ) = - \sqrt{n_k (\nu - n_k)}
        \Phi_{n_k - 1} ( y_k, \xi_k ) \delta (k , l),
$$
$$
a_{0 l} \Phi_{n_k}     ( y_k, \xi_k ) = s_k  
        \Phi_{n_k}     ( y_k, \xi_k) \delta (k , l),
$$
where $k$ and $l = 1, 2$. We close this section with a remarkable result: The
vibrational Hamiltonian $H_v$ can be written as
$$
H_v = - \hbar \Omega (a_{0 1}^2 + a_{0 2}^2)
$$
in terms of the energy operators $a_{0 1}$ and $a_{0 2}$. 

\section{The Jahn-Teller interaction}

The Jahn-Teller interaction Hamiltonian 
$H_{JT}$ for the studied system has the 
form \cite{2} 
\begin{equation}
H_{JT} = \kappa \hbar \Omega (\vecmu^+ \vecmu)^{(E)} (x_1 + x_2),
\label{hjtuv}
\end{equation} 
where 
${\vecmu^+}$ and ${\vecmu}$ represent the creation and
annihilation operators for the doubly degenerate electronic
states of the system 
(the label $E$ indicates that the operator acts on the electronic states)
and $\kappa$ is the strength of the 
Jahn-Teller coupling ($\kappa$ corresponds to $K$ in units of
$\hbar \Omega$). 

We want to write the operator $H_{JT}$ 
in terms of the operators $a_{\pm k}$ and $a_{0 k}$.
For this purpose, we start from the relations 
$$
{\partial\over \partial y_k} = - {1\over 2} \sum_{m = 0}^\infty
(2 a_{0 k})^m \left\{ 
{\rm e}^{+ {\rm i} \xi_k} a_{+ k} - (-1)^m
{\rm e}^{- {\rm i} \xi_k} a_{- k} + {\nu \over 2} [1+(-1)^m] \right\}
$$
and
$$
{1\over y_k} = - \sum_{m = 0}^\infty (2 a_{0 k})^{m -1} \left\{
{\rm e}^{+ {\rm i} \xi_k} a_{+ k} + (-1)^m 
{\rm e}^{- {\rm i} \xi_k} a_{- k} + {\nu \over 2} [1-(-1)^m] \right\}.
$$
The coordinate
$$
x_k = {1\over \alpha} ( \ln \nu - \ln y_k )
$$
can be expanded in the form of a series
$$
x_k = {\ln \nu \over \alpha}
    - {1       \over \alpha} \sum_{n = 1}^\infty 
    \frac{1}{n} \left( 1 - {1\over y_k} \right)^n 
$$
or more precisely
$$
x_k = {1\over \alpha} \left\{ \ln \nu -  \sum_{n = 1}^\infty 
\frac{1}{n} \left[
1 + \nu \sum_{m = 0}^\infty (2 a_{0 k})^{2 m} \right. \right.
$$
$$
+ \left. \left. \sum_{m = 0}^\infty (2 a_{0 k})^{m - 1} [
{\rm e}^{+ {\rm i} \xi_k} a_{+ k} + (-1)^m 
{\rm e}^{- {\rm i} \xi_k} a_{- k} ] \right]^n \right\}.
$$
As a result, the Hamiltonian $H_{JT}$ can be expressed as
$$
H_{JT} = {1\over \alpha} \kappa \hbar \Omega 
(\vecmu^+ \vecmu)^{(E)} \left\{ 2 \ln \nu
- \sum_{k = 1}^2 \sum_{n = 1}^\infty \frac{1}{n} \left[ 1 + \nu 
  \sum_{m = 0}^\infty (2 a_{0 k})^{2 m} \right. \right.
$$
\begin{equation}
+ \left. \left. \sum_{m = 0}^\infty (2 a_{0 k})^{m - 1} 
[ {\rm e}^{+ {\rm i} \xi_k} a_{+ k} + (-1)^m 
  {\rm e}^{- {\rm i} \xi_k} a_{- k} ] \right]^n \right\}.
\label{h_jtuv}
\end{equation}
The latter development in powers of the operators
$a_{\pm k}$ and $a_{0 k}$ (with k = 1,2) 
is of central importance for the treatment 
of the vibronic interaction. 

Following the approach in Ref.~\cite{2}, 
the wavefunctions of the electron-phonon system 
will be written as the overlap
\begin{equation}
| \beta n z \rangle = \int_0^{2 \pi} d\varphi 
| \beta     \rangle 
{\rm e}^{{\rm i} z \varphi}
{\rm e}^{{\rm i} \kappa a_+} (a_+ - \kappa)^n 
\Phi_0 ( y_1 , \xi_1 )
\Phi_0 ( y_2 , \xi_2 ),
\label{35}
\end{equation}
where $\beta$ refers to the lower branch ($\beta = l$)  or to the
upper branch ($\beta = u$) for which we have
\begin{equation}
| l \rangle = \cos {\varphi\over 2} |\theta \rangle - \sin {\varphi\over 2}
| \varepsilon \rangle,
\label{36}
\end{equation}
\begin{equation}
| u \rangle = \sin {\varphi\over 2} |\theta \rangle + \cos {\varphi\over 2} 
| \varepsilon \rangle,
\label{37}
\end{equation}
and where the creation operator $a_+$ is defined by
\begin{equation}
a_+ = a_{+ 1} \cos \varphi + a_{+ 2} \sin \varphi.
\label{38}
\end{equation}
The phase factor $\varphi$ in Eqs.~(\ref{35}-\ref{38}) 
is an arbitrary angle and the parameter $z$ in Eq.~(\ref{35}) 
characterizes the type of vibronic coupling. For example, the case 
$z = \frac{1}{2}$ corresponds 
to the states accessible by electric dipole 
radiation from the zero-phonon ground state.

The states $|\beta n z \rangle$ can be rewritten in a new form 
by expanding the right hand-side  of  (\ref{35}). This leads to
$$
| \beta n z \rangle = \sum_{p = 0}^\infty \sum_{q = 0}^n \int_0^{2 \pi}
d\varphi 
| \beta       \rangle {\rm e}^{{\rm i} z \varphi} 
{(-1)^q {\rm i}^p \kappa^{p+q} \over p!} C_n^q a_+ ^{n + p - q} 
|0 0          \rangle
$$
$$
=  \sum_{p = 0}^\infty \sum_{q = 0}^n (-1)^q { {\rm i}^p \kappa^{p+q} \over p!
}
C_n^q 
|\beta, n+p-q, z \rangle_0
$$
$$
=  \sum_{p = 0}^N (-{\rm i})^p
{n! \kappa^{p - n}\over p!}
L_n^{p - n} (\kappa^2) | \beta p z \rangle_0,
$$
where $C_n^q$ is a binomial coefficient,
$| \beta p z \rangle_0$ stands for 
$| \beta p z \rangle$ with $\kappa = 0$, and $L_n^{p-n}$ 
is a generalized Laguerre polynomial \cite{11}. Thus, 
the states $|\beta n 0 \rangle$ correspond to the case of 
a vanishing 
coupling and  $|\beta n \frac{1}{2} \rangle$ to the case of a weak
coupling. The states (\ref{35}) are eigenstates of the 
operators $b_{+k}$ (with $k= 1, 2$) and  of  
the Hamiltonian (\ref{hjtuv}). Therefore,  
they are appropriate for obtaining 
the energy due to the Jahn-Teller interaction.

To find the energy levels of the system with vibronic interactions, 
we calculate the matrix elements of the Hamiltonian $H_{JT}$
with respect to the states (\ref{35}) and use standard perturbation
theory. For this purpose, we use the notation 
$| n \rangle^0 = |\beta n z \rangle$ for the
state vectors in the zero-th order approximation and
consider the Jahn-Teller interaction as a perturbation of the system
described by $H_e + H_v$. The wavefunctions of the Hamiltonian 
(\ref{hjtuv}), in the first-order approximation of perturbation 
theory, have the expression
$$
|n \rangle ^1 = 
|n \rangle ^0 - 
\sum_{m = 0, m \ne n}^N 
{1 \over { E_0(m) - E_0(n) }} \, 
  {   {^0 \langle} m | H_{JT} |n \rangle ^0} \,
|m \rangle ^0.
$$

As a result, the first-order energy of the system is
$$
E = E_0 + E_{JT},
$$
where
$$
E_0    =    {^0 \langle n} | H_e + H_v | n \rangle ^0
$$
is the energy without the Jahn-Teller interaction and
$$
E_{JT} =    {^0 \langle n} | H_{JT}    | n \rangle ^0
$$
is the energy of the Jahn-Teller interaction. The calculation of
the wavefunctions and the energy corrections using the coherent
states (\ref{35}) is difficult because the expressions for the
matrix elements of $H_{JT}$ are very complicated. Moreover, the
utilization of the operator $\exp({\rm i} \kappa a_+)$ instead of the
unitary operator $\exp({\rm i} (\kappa a_+ - \kappa^\ast a))$,
where $a$ is the adjoint of $a_+$, 
in building the coherent states is rigorous only for the case of the 
harmonic oscillator corresponding to the ordinary Weyl-Heisenberg algebra.
For this reason, we calculate the matrix elements 
$E_{JT}$ by using the Jahn-Teller Hamiltonian 
(\ref{h_jtuv}). It can be proved that the matrix elements corresponding 
to the last term of Eq.~(\ref{h_jtuv}) vanish. As a final result, we obtain 
$$
E_{JT} = {1\over \alpha} \kappa \hbar \Omega 
\sum_{p, q = 0}^N (-1)^p {n!^2 \kappa^{p + q - 2n} \over p! q!} 
L_n^{p-n} ( \kappa^2 ) 
L_n^{q-n} ( \kappa^2 )
$$
$$
 \langle  \beta | (\vecmu^+ \vecmu)^{(E)} | \beta \rangle 
 \left\{ 2 \ln \nu - \sum_{k = 1}^2
\sum_{l = 0}^\infty \frac{1}{l} \left[ 1 + \nu \sum_{m = 0}^\infty 
(\nu - 1 - 2p)^{2 m} \right]^l \right\},
$$
where $N$ is such that  $N+1$  is the total number 
of states for the one-dimensional Morse oscillator.

\section{Conclusions}

We studied the vibronic coupling between doubly degenerate 
vibrations and doubly degenerate electronic states of an octahedral 
system. In order to describe the vibrations of the system, we used a new type
of
anharmonic coherent states.  We described the dynamical system by introducing 
two auxiliary variables
$\xi_k$ (with $k=1,2$) of extra-phase type and new eigenfunctions of 
the Hamiltonian for the  
anharmonic vibrations. The eigenfunctions of the Hamiltonian of the system
in absence of the vibronic interaction were builded using the overlap of
electronic
states and these new anharmonic coherent states. In the first-order 
approximation of
perturbation theory, the energies due to the Jahn-Teller interaction
were calculated. The results show that the anharmonic effects 
on the Jahn-Teller interaction can be expressed in a rigorous
form and are contained in the constants $s_k$ ($k=1,2$). The 
work reported in this paper
constitutes the most general
algebraic analysis of the 
Jahn-Teller interaction for $E \otimes \varepsilon$ 
octahedral systems with anharmonic vibrations.

\section*{Acknowledgements}
The authors are grateful to Prof.~E.~Duval (Universit\'e Claude Bernard 
Lyon~1) for interesting comments. Two of 
the authors (N.M.~A. and Gh.E.~D.) wish to
acknowledge the Institut de Physique
Nucl\' eaire de Lyon for the kind hospitality 
extended to them in the final stage of this paper 
and the CNCSIS (Romania)
for financial support under the grant 12625/1(1998).


\begin{thebibliography}{aa}

\bibitem{01} W. Moffit and W. Thorson, Phys. Rev. {\bf 108}, 1251 (1957).

\bibitem{02} U. \" Opik and M. H. L. Pryce, Proc. Roy. Soc. {\bf A238},
      425 (1957).
      
\bibitem{03} H. C. Longuet-Higgins, U. \" Opik, M. H. L. Pryce and H. Sack,
      Proc. Roy. Soc. {\bf A244}, 1 (1958).
      
\bibitem{04} H. C. Longuet-Higgins, Adv. Spectrosc. {\bf 2}, 429 (1961). 

\bibitem{05} I. B. Bersuker, Sov. Phys. - JETP {\bf 16}, 933 (1963). 

\bibitem{06} M. C. M. O'Brien, Proc. Roy. Soc. {\bf A281}, 393 (1964).

\bibitem{07} F. S. Ham, Phys. Rev. {\bf 166}, 307 (1968).

\bibitem{08} I. B. Bersuker and V. Z. Polinger, {\em Vibronic Interactions
        in Molecules and Crystals} (Springer Verlag, Berlin, 1989).
        
\bibitem{09} R. I. Badran and C. A. Bates, J. Phys. {\bf C3}, 6329 (1991).
        
\bibitem{010} M. D. Sturge, in {\em Solid State Physics}, eds. 
         F. Seitz, D. Turnbull and H. Ehrenreich, vol. 20 (Academic Press, 
         New York, 1967). 
          
\bibitem{011} I. B. Bersuker, {\em Electronic Structure and Properties
         of Transition Metal Compounds} (John Wiley and Sons, New York,
         1996).
         
\bibitem{012} R. Englman, {\em The Jahn-Teller Effect in Molecules and
Crystals}
         (John Wiley, London, 1972).

\bibitem{1} B. R. Judd, Can. J. Phys. {\bf 52}, 999 (1974).

\bibitem{2} B. R. Judd and E. E. Vogel, Phys. Rev. B {\bf 11},
    2427 (1975).
    
\bibitem{3} C. C. Chancey, J. Phys. {\bf A17}, 3183 (1984).

\bibitem{4} C. C. Chancey and B. R. Judd, J. Phys. {\bf A16}, 875
    (1983).
    
\bibitem{5} C. C. Chancey, J. Phys. {\bf A20}, 2753 (1987).

\bibitem{6} J. Rivera-Iratchet, M. A. de Or\' ue, M. L.
     Flores and E. E. Vogel, Phys. Rev. {\bf B47}, 10164 (1993).
     
\bibitem{7} M. C. M. O' Brien and C. C. Chancey, Am. J. Phys. 
     {\bf 61}, 688 (1993); H. G. Reik, J. Phys. {\bf A26}, 
     6549 (1993).
     
\bibitem{8} L. Landau and E. Lifshitz, {\em M\'ecanique Quantique}
    (Mir Publisher, Moscou, 1980); 
    R. G. Wybourne, {\em Classical Groups for 
    Physicists} (Wiley, New York, 1974);  
    A. E. Kondo and R. Truax, J. Math. Phys. {\bf 29}, 1396 (1988); 
    P. Cordero and 
    S. Hojman, Lett. Nuovo Cimento {\bf 4}, 1123 (1970).
    
\bibitem{9} C. C. Gerry, Phys. Rev. {\bf A33}, 2207 (1986); see also:
     C. C. Gerry, Phys. Rev. {\bf A31}, 2721 (1985).
     
\bibitem{10} N. M. Avram, Gh. E. Dr\v ag\v anescu and C. N. Avram,
     J. Optics {\bf B2}, 214 (2000);  Gh. E. Dr\v ag\v anescu and N. M.
     Avram, Zeitschr. Phys. Chem. {\bf 200 S}, 51 (1997).
     
\bibitem{11} M. Abramowitz and I. A. Stegun, 
             {\em Handbook of Mathematical Functions} 
             (Nat. Bur. Standards, Washington, 1964).
\end{thebibliography}
\end{document}